\pdfoutput=1
\documentclass[epj]{pw}
\usepackage[latin1]{inputenc}
\usepackage{amssymb}
\usepackage{graphics}
\usepackage{graphicx}
\usepackage{amsmath}

\usepackage[usenames,dvipsnames]{color}

\begin{document}
\title{Ultra--low noise differential AC-coupled photodetector for sensitive pulse detection applications}
\author{Patrick J. Windpassinger \thanks{\email{pwindpas@nbi.dk}}\inst{1,3} \and Marcin Kubasik\inst{2,4}  \and Marco Koschorreck\inst{2} %
\and Axel Boisen\inst{1} \and Niels Kj\ae rgaard\inst{1} \and Eugene~S.~Polzik\inst{1,2} \and J\"{o}rg Helge M\"uller \inst{1}% etc
}
\authorrunning{P. Windpassinger {\it et al.} }
\titlerunning{Ultra low noise  photodetector for pulse detection}
\institute{QUANTOP, Niels Bohr Institute, University of Copenhagen,
Blegdamsvej 17, 2100 Copenhagen, Denmark \and ICFO-Institut de
Ciencies Fotoniques, Mediterranean Technology Park, 08860
Castelldefels (Barcelona), Spain \and present address: Institut für
Laserphysik, Luruper Chaussee 149, 22761 Hamburg, Germany \and
present address: Clarendon Laboratory, Parks Road, Oxford OX1 3PU,
United Kingdom}
\date{Dated: March 19, 2009}
\abstract{We report on the performance of  ultra low noise
differential photodetectors especially designed  for  probing of
atomic ensembles with weak light pulses.  The working principle of
the detectors is described together with the analysis procedures
employed to extract the photon shot noise of light pulses with
$\sim1\,\mu$s duration. As opposed to frequency response peaked
detectors, our approach allows for broadband quantum noise
measurements. The equivalent noise charge (ENC) for two different
hardware approaches is evaluated to 280 and 340 electrons per pulse,
respectively which corresponds to a dark noise equivalent photon
number of $n_\mathrm{3dB}=0.8\cdot 10^5$ and
$n_\mathrm{3dB}=1.2\cdot 10^5$ in the two approaches. Finally, we
discuss the possibility of removing classical correlations in the
output signal caused by detector imperfection by using
double--correlated sampling methods.
}
\maketitle
\section{Introduction}\label{intro}
Many experiments in the  area of atomic physics and quantum
information science hinge on the possibility of detecting light
pulses of few microseconds duration with low photon numbers
($n_\mathrm{ph}\sim 10^5$) reliably, i.e., with ideally no noise
contribution from the detection electronics. For example, in the
specific case of interferometric measurements
\cite{Chaudhury2006}\cite{Usami2007}\cite{Echaniz2005}\cite{Windpassinger2007},
small differential signals in the interferometer outputs need to be
measured with quantum noise limited precision. Strong local
oscillators are in principle possible in some of these setups,
however, especially when interested in measuring broadband quantum
noise, low noise detectors greatly relax the requirements on the
technical noise of the local oscillator. The need for very low noise
analog photo detectors becomes even more apparent, when the number
of photons needs to be measured in a pulse with a photon flux too
high to employ direct photon counting techniques
\cite{Hilliard2008}. In general, any application where the number of
photons to probe a system is limited, e.g., to minimize the energy
deposited into a system in spectroscopy, will benefit from low
detection noise.

In the following we discuss the performance of photo-detectors
employed to faithfully measure the photon number difference of two
light beams. The difference in photon numbers of light pulses with
known shape and arrival time at the detector will contain the
information we wish to retrieve. The information should preferably
be extracted with a precision only limited by the intrinsic quantum
fluctuations of light -- light shot noise. Hence electronic and
classical noise contributions should be suppressed as much as
possible. To this end, we have developed differential photo
detectors based on two different commercially available front-end
hybrid amplifiers. In Version~I we use a similar layout as discussed
in \cite{Hansen2001}, based on Amptek\footnote{Amptek Inc., 14 De
Angelo Drive Bedford, MA. 01730 U.S.A.} amplifiers.  Version~IIis
based on a charge sensitive preamplifier and a pulse shaping module
from Cremat\footnote{Cremat Inc., 45 Union St. Watertown, MA 02472
U.S.A.}. The differences and performances are compared in table
\ref{tab1}.\\
\begin{table}[h]
\begin{center}
\begin{tabular}{l|p{2cm}|p{2cm}}\hline
    & Version I & Version II\\
    & AMPTEK    & Cremat \\
    \hline\hline
integrator & A250, external FET and feedback & CR 110, hybrid\\
\hline shaper & 2$\times$ A275, 3 pole, 330\,ns & CR-200-250\,ns, hybrid\\
\hline $n_{\mathrm{photon, 3 dB}}$ & $0.8\cdot 10^5$ & $1.2\cdot
10^5$\\
%\hline ENC & 280 $e^-$ in 1$\,\mu$s & 340 $e^-$ in
%1$\,\mu$s\\
\hline
\end{tabular}\caption{Comparison of the two detector versions.}\label{tab1}
\end{center}
\end{table}
These detectors were designed for the purpose of measuring quantum
noise in atomic ensembles \cite{Appel2008}, therefore both the
photon number impinging on the detector and the duration of the
light pulses is  restricted by the optical depth and the coherence
time of the ensemble \cite{Windpassinger2007} to the range of
$10^5-10^6$ photons per microsecond. We have built several units of
each type and with both approaches we consistently reach electronic
noise levels of ENC $ \sim 300$ for input pulse durations $\tau
\lesssim 1\,\mu$s.  We thus demonstrate an extension of the
technique proposed in \cite{Hansen2001} to the microsecond domain
(compared to picoseconds). Despite the longer pulses in our
measurements, we achieve an improvement of the noise performance by
a factor of 2-2.5 with respect to \cite{Hansen2001}. %%
%
%
%We have built several of these detectors and with both approaches we
%consistently reach an electronic noise level of ENC $ \sim 300$
%electrons per pulse for input signal durations $\tau \lesssim
%1\,\mu$s, which is am important improvement compared to
%\cite{Hansen2001}. In addition, we extend the working range of the
%detector for short pulses (ps range in \cite{Hansen2001}) to pulses
%in the microseconds rage and discuss the implications for the data
%analysis.
We start out by briefly reviewing the working principle of
the detectors. The data analysis procedures are discussed in detail
and the detector performance is evaluated.

\section{Technical background}
\begin{figure*}[t]\begin{center}
\includegraphics[width=.8\textwidth]{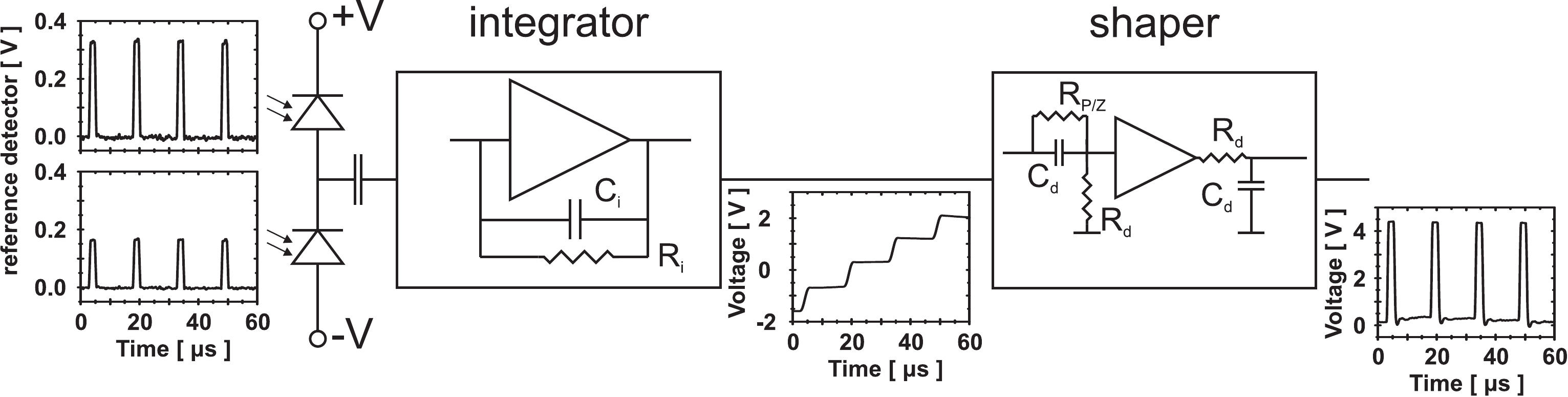}
\caption{Schematics of the detector electronics. When an imbalanced
signal is applied to the directly subtracted photodiodes as shown on
the left, the differential photoelectrons are integrated. The time
constant of the differentiator is adjusted such that it matches the
decay time of the integrator. In consequence the signal is zero
between the pulses and the integral of the output pulses after the
shaper is proportional to the differential photoelectrons.}
\label{diag}
\end{center}\end{figure*}

\begin{figure}[h]\begin{center}
\includegraphics[width=.75\columnwidth]{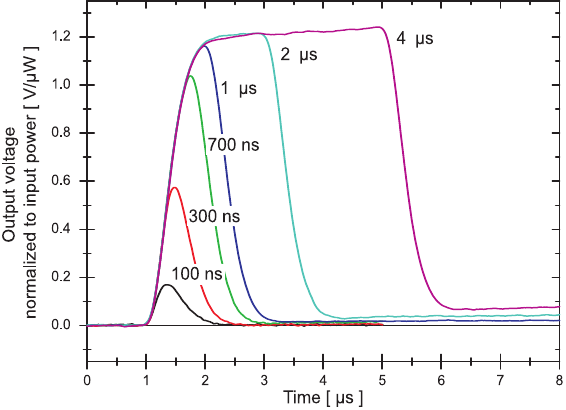}
\end{center}
\caption{Output signal samples for different input pulse durations.
The output signal has been normalized to the power of the input
beam. The duration of the electronic output signal is extended by
twice the shaping time. For long pulses, i.e. high photon numbers
per pulse,  a clear pulling of the detector baseline can be observed
as an output signal which does not coincide with the level prior to
the pulse.} \label{blpull}
\end{figure}

A schematic block diagram of the detector circuits we are
considering is shown in Fig. \ref{diag}. The current through
two biased PIN photo diodes (Hamamatsu S3883) is directly
subtracted and the difference current is
AC coupled into a charge sensitive amplifier. The node between the
photodiodes is in some implementations connected with high
resistance (50\,M$\Omega$ -- 1\,G$\Omega$) to the midpoint of the
supply voltages in order to avoid large difference in bias voltage
for diodes with differing dark currents. In Version~I, the charge
integrator is built with a discrete external FET input stage coupled
to a high-speed operational amplifier (AMPTEK A250) using external
R-C--feedback.  In Version~II a self contained unit (Cremat CR110)
with similar functionality but fixed integration gain is used. The
integrated signal is buffered and derived with Gaussian filters
(shaping amplifiers), where the shaping time of Version~I is user
defined.  Version~II features an integrated shaper circuit module
with fixed shaping time (Cremat CR200) of $250\,$ns is used. Typical
signals at different stages in the detector for a $1\,{\mu}$s
unbalanced input light pulses are shown as insets in figure
\ref{diag}. The high gain-bandwidth product and fast slew rate of
the amplifiers in the integrator circuit sets the minimum rise time
at the integrator output to $\sim 5\,$ns. A pulse with a duration of
more than the rise time will be transformed into a linearly rising
voltage at the output of the integrator, where the slope is
proportional to the differential photo current and the total step
size proportional to the total (differential) charge. The
discharging time $\tau_i=C_iR_i$ of the integrator feedback circuit
is chosen such that it has little influence during the actual light
pulse, $\tau_i \gg \tau$. \\
Deriving the integrated  signal with a Gaussian filter, in the
simplest form realized by an C-R--R-C combination as indicated in
figure \ref{diag}, will result in a pulse with an area proportional
to the integrated charge. For input pulses considerably shorter than
the shaping time of the filters, the duration of the output pulse
will be approximately 2.4 the shaping time with now both pulse
height and pulse area proportional to the integrated charge. This is
the usual mode of operation for charge sensitive front-ends in X-ray
and particle detectors. An initial light pulse with a duration
longer than the shaping time  will result in an output signal which
is widened approximately by twice the shaping time. In our
particular implementation the shaping time of Version~I is set to
330\,ns, thus a light pulse of $\tau\approx100$\,ns duration will
result in an electronic pulse of $\sigma\approx 800\,$ns. Figure
\ref{blpull} shows output pulse samples for unbalanced rectangular
input light pulses of various durations incident on the detector. It
should be clear  that  the type of integrating
detector presented here is not suited to provide information about
the temporal shape of the input signal on a time scale shorter than
the shaping time. The choice of a specific shaping time is a
compromise between the desired time resolution and the digital
signal sampling bandwidth for postprocessing of the data. The input
referred electronic noise level in the chosen amplifier
configuration depends essentially on the combined size of detector
capacitance and the gate capacitance of the input FET
\cite{CSA-Noise}. In addition there is a dependence of the noise on
shaping time and the digital postprocessing. In the remainder of the
article we will characterize the output
referred noise level in different operating regimes. \\

In the simple Gaussian filter configuration, an active C-R--R-C high
-- low pass combination, the output level of the electronic signal
does restore to zero only on the timescale of the integrator
discharge time $\tau_i$. More formally, the transfer function of the
combined system has a pole at $\tau_i$ which leads to a non--zero
signal after the pulse. The pole in the transfer function of the
integrator  can be compensated by adding a zero in the transfer
function of the subsequent filter. This is achieved by adding the
resistor $R_\mathrm{P/Z}$ shown in Fig. \ref{diag}.  Adapting the
resistor value to the preceding  integrator allows to reduce the
effect of the pole.  In practice, the cancelation is never perfect
leading to weak but long tails (baseline pulling or pulse pile-up)
in the detector response. This imperfection is evident in Fig.
\ref{blpull} for pulses with durations longer than $1\,\mu$s. It has
to be accounted for properly in the analysis since it is
proportional to the input pulse photon number and thus can enter as
a classical noise contribution (autocorrelation) in a statistical
analysis of pulse areas from train of pulses. Under ideal conditions
the average detector signal is balanced to zero at all times and the
noise is calculated from fluctuations around the zero level. In
practice, drifts in the optical setup can lead to slowly varying
signal imbalance over time and cause considerable contributions for
baseline pileup.
\\
To analyze the output signal of the detectors we use a digital
storage oscilloscope (Agilent Infiniium 54832D) with analog
bandwidth limitation to 20\,MHz to avoid under-sampling and folding
of RF interference. The signal trace is digitized and stored for
numerical postprocessing. By treating the electronic signal in this
way, the only additional noise contribution after the detector
front-end originates  from the input channel noise of the
oscilloscope and the digitization noise of the 8\,bit A/D converter.
With high gain in the front-end and by choosing equipment of
suitable quality, these noise sources can be neglected.

\section{Noise analysis and detector performance}
%\subsection{Balanced two beam setup}
%To test the performance of the detectors we use the setup shown in
%figure \ref{setup}. A light beam derived from an external grating
%stabilized, frequency stabilized  diode laser at 780\,nm, pulsed
%with a standard acousto--optical modulator,  is coupled into a mode
%leaning fiber. At the output of the fiber the polarization is
%cleaned with a polarizing beam splitter (PBS) and a large fraction
%s split--off with a second PBS and directed onto fast photon
%etector for power monitoring.  The actual probe beam is after
%focussed onto one photo diode of the differential detector. The
%power calibration and data acquisition procedure is fully automated
%via a computer interface which controls both the storage
%oscilloscope and a motorized wave plate used to adapt the beam
%powers.
%\begin{figure}[h]\begin{center}
%\includegraphics[width=.95\columnwidth]{fig3_setup2.pdf}
%\caption{Test setup. The light beam derived from a grating
%stabilized diode laser is mode cleaned in a fiber. A large fraction
%of the power is diverted into a fast, calibrated reference detector
%and the rest of the beam is further attenuated, split into two parts
%and focussed onto the two photodiodes. The output is directly
%digitized on a storage oscilloscope and saved to a hard drive for
%postprocessing. } \label{setup}
%\end{center}\end{figure}

\subsection{CW noise spectra and time domain integration}
We model the light pulses incident on the detector as coherent state
excitations with a temporal mode function given by the rectangular
pulse envelope. Denoting the coherent light state as $|\alpha
\rangle$, one has a mean photon number $\langle \alpha| \hat{N} |
\alpha \rangle = \langle \alpha| a^\dagger a | \alpha \rangle=
\bar{N} $ in every pulse and intrinsic Poissonian fluctuations of
size $\langle(\delta \hat{N})^2\rangle = \langle \alpha|
(\hat{N}-\bar{N})^2 | \alpha \rangle = \bar{N}$ from pulse to pulse
\cite{Scully1997}. Thus the expected noise (variance) of a
measurement is equal to the mean number of photons in the pulse.
This picture remains valid when a single coherent state with
$\bar{N}$ photons is split into two $\bar{N}_1+\bar{N}_2=\bar{N}$ on
a beam splitter and both coherent states are measured individually.
The variance of the combined output signal -- e.g. balanced
difference detection of the two states as in our case -- will still
be $\langle(\delta (\hat{N_1} - \hat{ N_2}))^2\rangle = \bar{N}$. As
a typical feature of quantum noise, the variance of the output
signal of the balanced detector should thus scale linearly with the
mean photon number per pulse.

To describe  the  noise properties of the detector we consider the
following block diagram:\begin{center}
\includegraphics[width=.65\columnwidth]{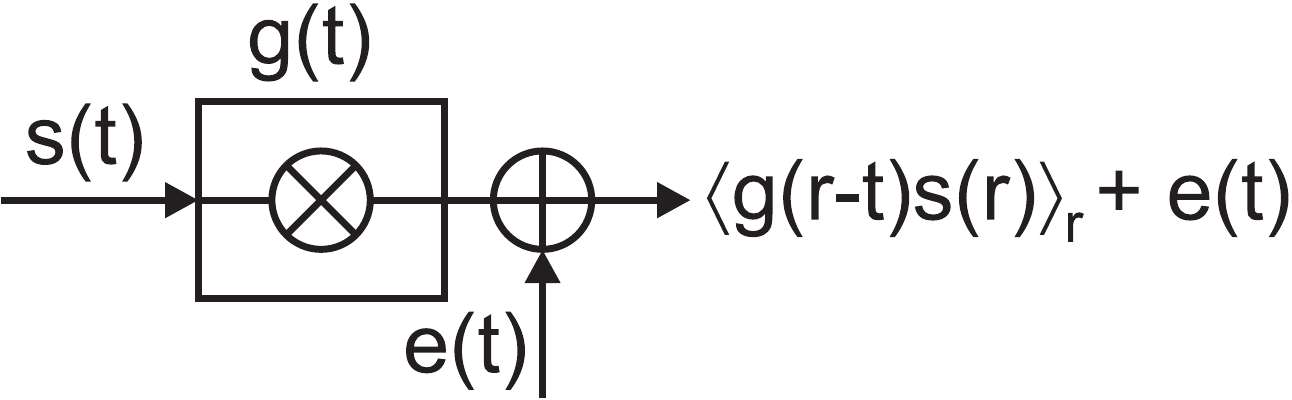}\end{center}
The input signal is characterized by the random variable $s(t)$. The
detector gain $g(t)$ only acts on  the input signal and the output
corresponds to the convolution of the two processes, $o(t)=\langle
g(r-t)s(r)\rangle_r$. We add a random electronic signal $e(t)$ at
the output to account for the dark detector noise. The term noise is
usually connected to the autocorrelation function $\Omega(t)=\langle
o(r)o(r-t)\rangle_r$ of a time dependent random signal $o(t)$, or
the Fourier transform of the autocorrelation function
$\Omega(\omega)$, which is the so--called spectral noise power
density. According to the Wiener--Khinchin--theorem,
$\Omega(\omega)=\langle|o(\omega)|^2\rangle$, where $o(\omega)$ is
the Fourier transform of $o(t)$ \cite{Reif1987}. Consequently, when
invoking Parseval's theorem, the spectral noise power density which
can conveniently be measured on a spectrum analyzer  can be
converted into:
\begin{equation}
    \Omega(\omega)=|g(\omega)|^2 |s(\omega)|^2 + |e(\omega)|^2
\end{equation}

Modeling a primary detection event as a {$\delta$}-function current
spike  at the input implies a flat spectral noise power density
$|s(\omega)|^2=s_0$ (white noise) with a strength directly
proportional to the incident photon flux. This flat spectral noise
density is filtered and amplified by the detector response
$g(\omega)$ (including transit time effects in the photo diodes).
Due to its frequency dependent transimpedance gain $g(\omega)$, the
detection electronics allows one to measure signals only within a
certain bandwidth. The complex gain characteristics can in principle
be extracted directly from the Fourier transform of the pulse
response of the detector. When a continuous coherent light beam is
applied in a balanced way to the detector, the input noise of light
has a flat noise distribution, thus the observed spectral noise
density of the electronic output signal can be used directly to
determine the modulus of the transimpedance and the detector
electronic base noise:
\begin{equation}
    |g(\omega)|^2=\frac{\Omega(\omega) - |e(\omega)|^2}{s_0}
\end{equation}

\begin{figure}[th]\begin{center}
\includegraphics[width=\columnwidth]{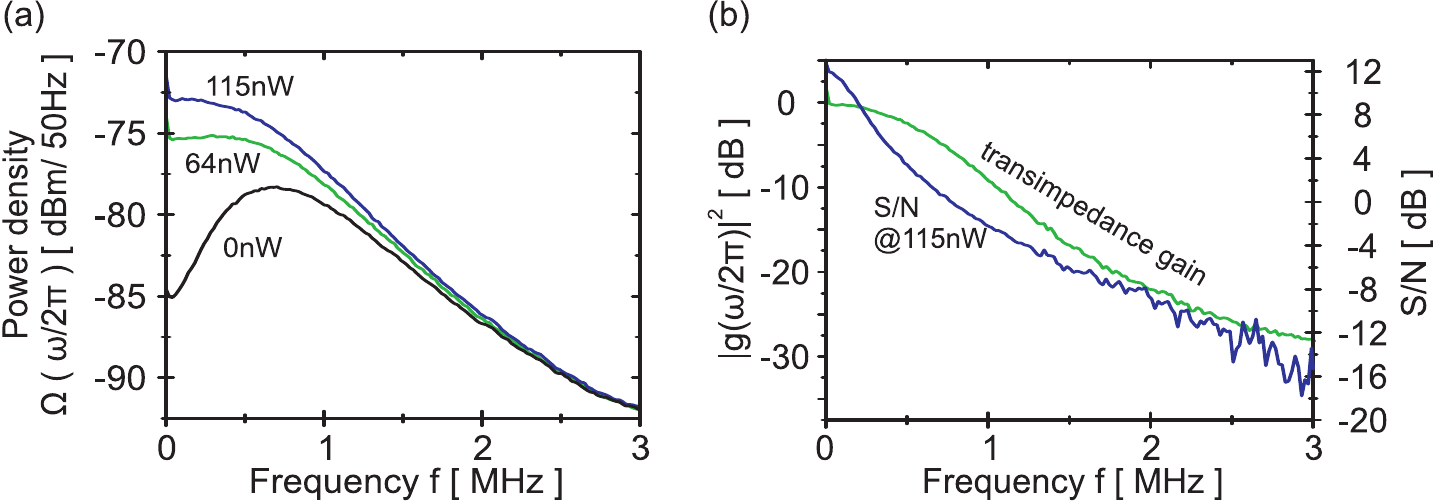}
\caption{(a) Power spectrum of the detector when different, balanced
dc light power levels are applied to the detector. (b) Frequency
response of the detector, calculated from the raw data trace in (a)
at 115\,nW dc light power level. The gain drops by 3\,dB within
$\sim 600$\,kHz, which corresponds well to a shaping time of $\sim
1.5\,\mu$s.} \label{spectrum}
\end{center}\end{figure}

In Fig. \ref{spectrum}(a) we show the noise power density
$\Omega(\omega/2 \pi )$ for Version~I of the detector when
illuminated cw with balanced light beams at different power levels.
We subtract the electronic noise from the data with a known light
level incident on the diodes and extract the transimpedance gain
$|g(\omega/2\pi)|^2$ of the detector. From Fig. \ref{spectrum}(b) we
observe that the gain drops by 3\,dB at a frequency somewhat below
600\,kHz. We also show the ratio of signal noise to electronic noise
for input dc light powers of $P=115$\,nW and observe a ratio of
12\,dB at low frequencies which drops towards higher frequencies.

\begin{figure}[th]\begin{center}
\includegraphics[width=\columnwidth]{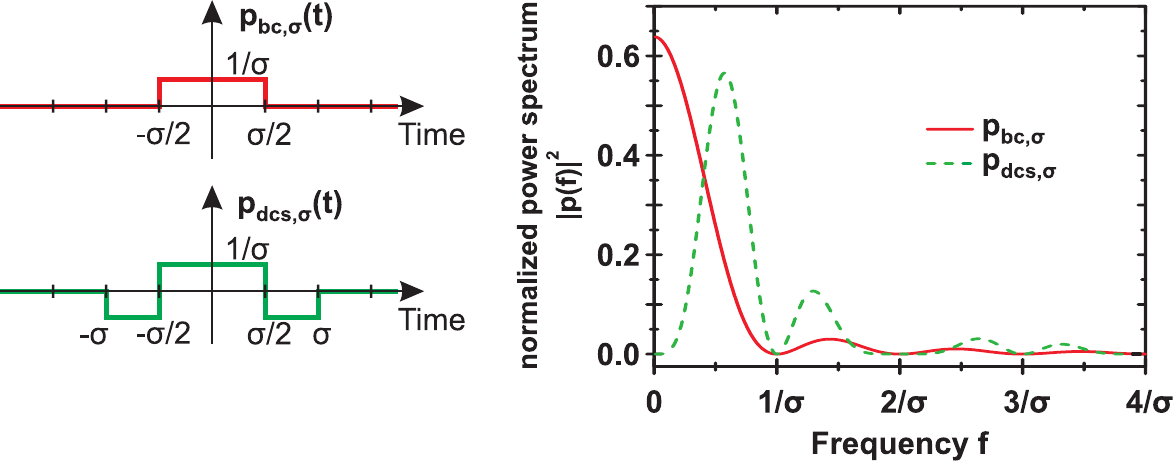}
\caption{Integration pulse shapes $p_{\mathrm{bc},\sigma}(t)$ and
$p_{\mathrm{dcs},\sigma}(t)$ and their power spectra
$|p_{\mathrm{bc},\sigma}(\omega/2\pi)|^2$ and
$|p_{\mathrm{dcs},\sigma}(\omega/2\pi)|^2$. The integrated power
spectra have been normalized to unity on the interval $[0,\infty]$.}
\label{theo_spectrum}
\end{center}\end{figure}

To treat the case of pulsed illumination in the frequency domain,
the recorded cw-spectrum for white noise input has to be multiplied
with the spectrum of the light pulse shape. For a train of
independent square pulses $p_{\mathrm{bc},\sigma}(t)$ with duration
$\sigma$:
\begin{equation}
\label{p_tau} p_{\mathrm{bc},\sigma}(t)=\frac{\Theta(t + \sigma/2) -
\Theta(t - \sigma/2)}{ \sigma},
\end{equation}
where $%\begin{equation}
\Theta(x)=\left\{
\begin{array}{ccc}
1 & \mbox{for} & x>0 \\
0 & \mbox{for} & x \leq 0
\end{array}
\right. $%\end{equation}
is the Heaviside step function, the power spectrum  is:
\begin{equation}\label{pulsespec}
    |p_{\mathrm{bc},\sigma}(\omega)|^2=\left( \frac{\sin( \omega \sigma/2)}{\omega
    \sigma/2}\right)^2
\end{equation}
Assuming now the simple boxcar integration window of duration
$\sigma$ given by equation (\ref{p_tau}) to determine the pulse area
we get the total detected noise power:
\begin{eqnarray}\label{timedomain}
     P_n         &=& \int_0^\infty \Omega(\omega) |p(\omega)|^2 d\omega\\
 \nonumber    &=& \int_0^\infty |g(\omega)|^2 s_0 |p(\omega)|^2 d\omega + \int_0^\infty |e(\omega)|^2 |p(\omega)|^2 d\omega
\end{eqnarray}
Both the temporal shape and the spectrum of the pulse are
illustrated in Fig. \ref{theo_spectrum}. Experimentally the noise
power is determined from the variance of output pulse areas of a
large number of independent pulses. The real expression taking the
separation between pulses into account reduces to (\ref{timedomain})
in the limit of high number of pulses \cite{Barnes1971}. By using
measurements with a spectrum analyzer we can thus predict the noise
scaling expected when using real pulses. The above expression can be
used for pulse durations considerably longer than the shaping time.
With straightforward modifications also the case of integration
windows different from the input pulse length can be treated.\\

From the spectrum of the boxcar window eq.(\ref{pulsespec}) it is
clear that especially low frequency electronic noise, such as slow
baseline drifts but also the baseline pileup as discussed above will
contribute significantly to the total output noise. To circumvent
this problem, we change from the boxcar integration window
$p_{\mathrm{bc},\sigma}(t)$  to a balanced, double--correlated
sampling (dcs) function:
\begin{equation}
p_{\mathrm{dcs},\sigma}(t)= \frac{ \Theta(t + \sigma/2) - \Theta(t -
\sigma/2)}{\sigma/2} - \frac{ \Theta(t + \sigma) - \Theta(t -
\sigma)}{ \sigma}
\end{equation}
The resulting power spectrum:
\begin{equation}\label{dcpulsespec}
    |p_{\mathrm{dcs},\sigma}(\omega)|^2=4 \left(  \frac{\sin( \omega \sigma/2)}{\omega
    \sigma/2} -  \frac{\sin( \omega \sigma)}{\omega
    \sigma}\right)^2
\end{equation}
has no contribution at $\omega=0$. The pulse shape
$p_{\mathrm{dcs},\sigma}(t)$ and its power spectra is also
illustrated in Fig. \ref{theo_spectrum}.
Using an integration function as $p_{\mathrm{dcs},\sigma}(t)$ thus
reduces the influence of low frequency noise contributions and
allows one to cancel the effect of baseline pulling. Generally, the
integration with a certain gating function $p(t)$ can be understood
as frequency (band) pass filtering of the signal with the power
spectrum of the gating function $|p(\omega)|^2$. In the particular
case of $p_\mathrm{dcs}$, the window function is similar to the one
used for the  double--correlated sampling which is routinely applied
in CCD camera  readout units in order to reduce correlated noise
sources.

\subsection{Noise in pulsed operation}
In a typical experimental application  we use $k$ light pulses $p_i$
of duration $\tau$ of the order of some microseconds and repetition
period $r$, of the order of some tens of microseconds
\cite{Appel2008} (compare figure \ref{diag}).
%A representative detector output trace originating
%from 50 pulses of $\tau=2\,\mu$s is shown in Fig. \ref{pulseint}.
%\begin{figure}[ht]\begin{center}
%\includegraphics[width=.8\columnwidth]{fig7_20070215_ac_balanced_pulses.pdf}
%\caption{Electronic output signal as recorded on the oscilloscope
%for 50 light pulses of $\tau=2\,\mu$s duration, plotted on top of
%each other. The light noise can be clearly distinguished from the
%electronic noise level. The signal is integrated with a gating
%functions as discussed in the text.  } \label{pulseint}
%\end{center}\end{figure}
The electronic differential detector signal $S(t)$ is acquired on an
storage oscilloscope  and integrated with the gating function
$p_{\mathrm{bc},\sigma}(t)$ or $p_{\mathrm{dcs},\sigma}(t)$ to give
the pulse area normalized to its duration. Time delays, e.g. from
the response time from the pulsing device are taken into account by
a time translation $t \to t-t_0$ of the integration window. In
general, both the duration of the integration window $\sigma$ and
its position in time have to be optimized for each
detector/experimental setup. After the integration, we are left with
the pulses $p_i$:
\begin{equation}\label{integal}
    p_i= \int_{-\infty}^{+\infty} p_{\mathrm{bc},\sigma}(t) S(t) dt =
    \frac{1}{\sigma} \int_{t_0 + (i-1) r}^{t_0 + (i-1) r + \sigma}
    S(t) dt
\end{equation}
of which we calculate the variance\\
$%\begin{equation}\label{variance}
    \delta^2 p \equiv \frac{1}{k}\sum_{i=1}^k \left(p_i^2 -
    \left(\frac{1}{k}\sum_{i=1}^k p_i\right)^2\right)
$%\end{equation}
to evaluate the noise of the signal.

\subsubsection{Baseline subtraction}
When an imbalanced signal is being measured, the baseline pulling
encountered  in Fig. \ref{blpull} has to be taken into account. When
we use the double--correlated sampling gating function
$p_{\mathrm{dcs},\sigma}(t)$ we obtain:
\begin{eqnarray}\label{baseline}
p'_i &=&  \int_{-\infty}^{+\infty} p_\mathrm{dcs}(t) S(t) dt \\
\nonumber    &=& \frac{1}{\sigma} \left( \int_{t_0 + (i-1) r}^{t_0 +
(i-1) r + \sigma} S(t) dt \right) \\
\nonumber &-& \frac{1}{\sigma} \left(\int_{t_0 + (i-1) r -
\sigma/2}^{t_0 + (i-1) r}
    S(t) dt + \int_{t_0 + (i-1) r + \sigma}^{t_0 + (i-1) r + 3\sigma/2 }
    S(t) dt\right)\\
\nonumber &\equiv& p_i - b_i
\end{eqnarray}
It is clear that integrating  $S(t)$ with
$p_{\mathrm{dcs},\sigma}(t)$, i.e., also before and after acts as to
subtract the mean of two signal baseline intervals before and after
the light pulse. Since, in order to extract an estimate for the
baseline level, the output signal of the detector is in total
integrated over a longer time $2\sigma$ than for the simple boxcar
window function, the contribution of high frequency  electronic
noise to the total noise \textit{energy} of a pulse area signal is
increased. Depending on the detailed behavior of the low frequency
noise spectrum and under the constraints given by the desired pulse
repetition rate in an experiment, the shape and duration of the
baseline sampling interval can be optimized to minimize the
additional contribution of the electronic noise. The
double--correlated sampling discussed can be understood as taking
the two sample variance on the dark detector electronic signal, thus
removing correlations in the bare detector signal \cite{Allan1966}.

%\subsubsection{Two sample variances}
%Correcting for the detector baseline pulse pileup  only helps to
%reduce the correlations between pulse area signals introduced by
%imperfections in the detector electronics. In an experiment
%classical noise is also introduced by instabilities in the setup,
%e.g. acoustic vibrations or thermal drifts with dominating
%frequencies between the inverse of the pulse repetition period and
%the inverse of the duration of a long pulse train. Such fluctuations
%can be tackled by subtracting  two consecutive pulses,
%$p'_i-p'_{i+1}\to p'_i$ closer in time than the correlation time of
%the environmental noise. Determining the noise of differences of
%adjacent pulses amounts to evaluating the first Allan variance
%\cite{Allan1966}. The double--correlated sampling discussed above
%can be understood as taking the two sample variance on the dark
%detector electronic signal, thus removing correlations in the bare
%detector signal.

\section{Detector performance}
\begin{figure}[h]\begin{center}
\includegraphics[width=.8\columnwidth]{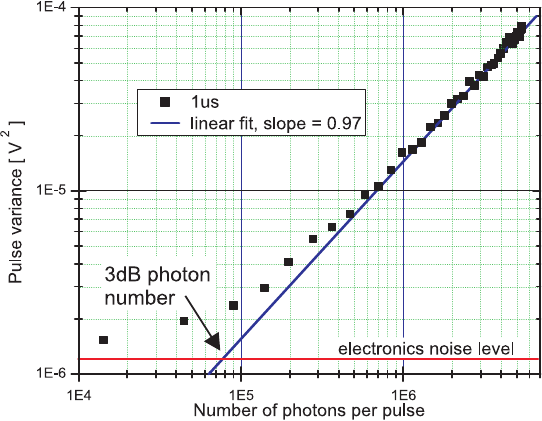}
\caption{Pulse variance for different pulse photon numbers. The data
has been analyzed using the simple boxcar gating function
$p_{\mathrm{bc},\sigma}$ with a $\sigma=1.25\,\mu$s integration
window around a $\tau=1\,\mu$s initial light pulse. The slope of the
linear fit in the log-log plot to the light noise dominated part  is
one, confirming that the observed noise is light shot noise. The
3\,dB photon number $n_\mathrm{3dB}=8\cdot10^4$ corresponds to a rms
electronic noise of ENC=280 electrons in the integration window. }
\label{SNdata}
\end{center}\end{figure}

Within the  framework presented above, we now analyze the
performance of the different detector versions. As a figure of merit
we consider the electronic noise equivalent light shot noise, i.e.,
the number of photons per light pulse required to generate the same
noise as the detection electronics has intrinsically ($3\,$dB
level). This photon number can then be converted into the equivalent
noise charge (ENC) which is just the corresponding standard
deviation. To determine this level we send balanced pulse trains of
several hundred light  pulses onto the detector, evaluate the
variance of the integrated signal and plot the variance as a
function of the mean photon number in each pulse. The result for
$\tau=1\,\mu$s pulse duration is shown in Fig. \ref{SNdata}. Here we
have used the simple integration gating function
$p_{\mathrm{bc},\sigma}(t)$. The scaling of the pulse variance data
with mean photon number per pulse in a log-log plot shows a slope of
one which confirms that the observed noise is due to light shot
noise. The optimal integration window $p_{\mathrm{bc},\sigma}(t)$ is
found by optimizing the $3\,$dB photon number with respect to  the
window duration $\sigma$.

\begin{figure}[th]\begin{center}
\includegraphics[width=.8\columnwidth]{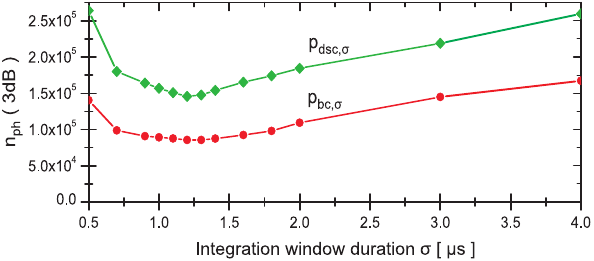}
\caption{By varying  the duration of the integration window and its
position with respect to the pulse, the optimal integration
parameters can be determined. When using the simple boxcar window
function we find the optimal integration duration at
$\sigma=1.25\,\mu$s with a 3\,dB photon number of
$n_\mathrm{3dB}\sim 0.8\cdot10^5$. When implementing baseline
subtraction via the gating function $p_{\mathrm{dcs},\sigma}(t)$,
the  3\,dB noise level rises by a factor of two,
$n_\mathrm{3dB}\sim1.6\cdot10^5$. The additional electronic noise
contribution when using $p_{\mathrm{dcs},\sigma}(t)$ is due to the
increased total signal integration time. } \label{windowopti}
\end{center}\end{figure}

In Fig. \ref{windowopti} the optimization procedure is illustrated
for both integration gating functions $p_{\mathrm{bc},\sigma}(t)$
and $p_{\mathrm{dcs},\sigma}(t)$. Evidently, the optimum duration of
the integration window for short pulses is governed by the shaping
time.
When the pulse integration window is too short, the electronic noise
is very large compared to the light noise in the time window, thus
the 3\,dB level is rather high. When the integration window is much
longer that the actual light pulse, the ratio of light shot noise to
electronic noise is artificially decreased and therefore the $3$\,dB
level rises again. The optimal value is found at approximately  the
initial pulse duration plus twice the shaping time. The optimal
noise data for Version~I of the detector  is $n_\mathrm{3dB}=8 \cdot
10^4$  electronic noise equivalent photon number for pulses shorter
than the shaping time, i.e., durations $<1.5\,\mu$s. Version~II
performs almost equally well with $n_\mathrm{3dB}= 1.2\cdot 10^5$
photons per microsecond. With  quantum efficiency of the employed
photodiodes exceeding  90\%, the photon numbers can be converted to
the more familiar  equivalent noise charge yielding ENC=280
electrons for version~I and ENC=340 electrons for version~II  when
applying light  pulses of up to $\tau\lessapprox 1\,\mu$s duration.
Both values are very close to those expected for the employed
front-end amplifier components. We attribute this improvement
compared to \cite{Hansen2001} to a very careful choice of the
electric components and a careful circuit layout together with the
analysis presented here. Specifically, photon numbers of $n\gtrsim
10^5$ can be reliably detected at the shot-noise level. The 3\,dB
levels  have been confirmed by an analysis of  cw noise power
density spectra whereof examples have been shown in Fig.
\ref{spectrum}. By analyzing these according to formula
(\ref{timedomain}), the expected noise performance in pulsed
operation can be inferred and an agreement between the values
obtained from both approaches confirms that, e.g., transient effects
due to light pulse switching do not affect the detector performance.
\\

\begin{figure}[ht]\begin{center}
\includegraphics[width=.8\columnwidth]{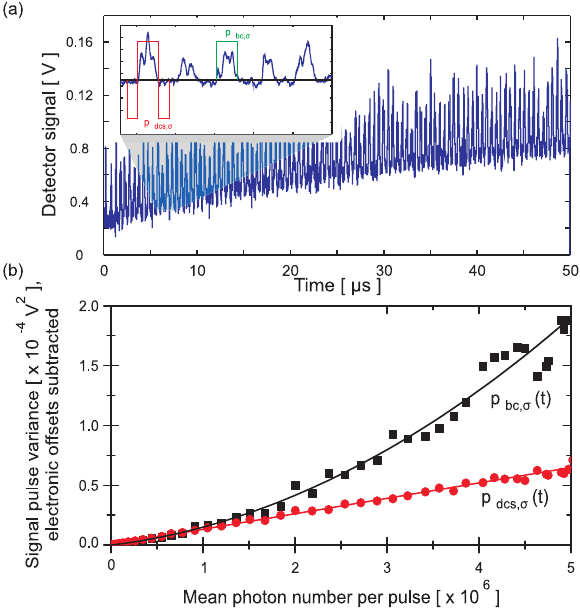}
\caption{(a) Due to technical imperfections in the pole/zero
cancelation of the Gaussian filter, the detector baseline piles up
when the incident light pulses are not fully balanced. (b) When
analyzing the signal from an imbalanced input, the variance of
pulses integrated with the simple integration gating function
$p_{\mathrm{bc}}$ shows a clear contribution of classical --
correlated noise. Using the baseline subtracting window function
$p_{\mathrm{dcs}}$, the linear scaling characteristic for shot noise
is regained. In both data sets, the electronic noise level has been
subtracted.} \label{20070301_dc_sampling}
\end{center}\end{figure}

As discussed above, the balancing of the detector may change during
the measurement and the resulting electronic signal pileup may make
a baseline subtraction by using the double--correlated sampling
window function $p_{\mathrm{dcs},\sigma}$ necessary. Fig.
\ref{20070301_dc_sampling}(a) shows the effect of baseline pileup
for a train of slightly imbalanced input signals. The
noise scaling results for the two gating  functions
$p_{\mathrm{bc},\sigma}(t)$ and $p_{\mathrm{dcs},\sigma}(t)$ are
compared in Fig. \ref{20070301_dc_sampling}(b). When applying the
simple integration function $p_{\mathrm{bc},\sigma}(t)$, a clear
quadratic (classical) noise contribution can be observed. In
contrast, when  the double--correlated sampling
$p_{\mathrm{dcs},\sigma}(t)$ is employed, the linear scaling and
thus shot noise limited performance is observed.  The linear part of
the data obtained with $p_{\mathrm{bc},\sigma}(t)$, i.e, its light
shot noise contribution corresponds exactly to
the light shot noise level extracted with $p_{\mathrm{dcs},\sigma}(t)$. \\
The price paid for using double--correlated sampling is that one
unit of electronic noise is added. By comparing the 3\,dB photon
number level in figure \ref{SNdata}, $n_\mathrm{3\,dB}=0.8 \cdot
10^5$ obtained with $p_{\mathrm{bc},\sigma}(t)$ for a balanced
signal with the level $n_\mathrm{3\,dB}=1.6 \cdot 10^5$ extracted in
figure \ref{windowopti} with double--correlated sampling window, we
observe this additional electronic noise contribution from the
analysis. In our typical experiments we use photon numbers per pulse
$n \approx 10^7$ which renders the electronic noise negligible
\cite{Appel2008}. We also implement the double--correlated sampling
integration function in order to compensate for detector
imperfections. The best performance of the detector is, of course,
reached when baseline subtraction is not necessary, i.e., when the
detector is balanced all the time. For very hight photon numbers,
e.g. $n>10^8$ per pulse, the shot noise alone can create an
imbalance which in turn may pull the baseline significantly.
Therefore, in this case, baseline subtraction is compulsory. On the
other hand, in this regime the  electronic noise is much lower than
the shot noise of light, and consequently, additional unit of
electronic noise is of little importance.

\section{Conclusion}
We have presented the working principle of ultra low noise
differential integrating AC photo detectors and discussed two
different hardware realizations with components from different
suppliers. The cw and pulsed noise performance of both detectors has
been discussed in detail with special emphasis on the analysis
procedures for pulsed operations. We have demonstrated that the
ultra--low noise performance of our detectors extends into the range
of microsecond long pulses which is of considerable importance for
atomic physics applications \cite{Appel2008}. There, duration and
photon number per pulse is set by the desired coupling strength
between the light pulses and the atomic ensemble  and is typically
in the range of $10^5-10^6$ photons distribute over few
microseconds. In this regime the noise performance of the detectors
discussed here is considerably better than what has been previously
reported \cite{Hansen2001}. %%
%%
%
%We have shown that both detector versions perform  better than the
%one presented in \cite{Hansen2001}, exhibiting an  rms equivalent
%noise charge of ENC $<330$ electrons.  Especially, we have shown
%that the ultra low noise performance of our detectors can be
%extended into the range of pulse durations of several microseconds,
%which is of considerable importance for atomic physics applications.
The influence of correlated noise sources  has been discussed, for
example classical pulse correlations generated by the pulling of the
detector baseline for imbalanced detector operation. A baseline
subtraction scheme has been proposed and successfully used to
circumvent these shortcomings. Correction for the baseline pulling
is generally only necessary when rather high photon numbers are
considered. In this case, the additional electronic noise due to the
analysis procedure is negligible compared to the detected light shot
noise.

\section*{Acknowledgements}
This work was funded by the Danish National Research Foundation, as
well as the EU grants QAP and EMALI.\\


\begin{thebibliography}{13}

\bibitem{Chaudhury2006}
Chaudhury~S, Smith~G~A, Schulz~K, Jessen~P~S 2006
  Continuous nondemolition measurement of the Cs clock transition pseudospin
\textit{Phys. Rev. Lett.} \textbf{96}
  043001

\bibitem{Usami2007}
Usami~K, Kozuma~M 2007
 Observation of a topological and parity-dependent phase of m=0 spin states
 \textit{Phys. Rev. Lett.} \textbf{99} 140404

\bibitem{Echaniz2005}
de~Echaniz~S~R, Mitchell~M~W, Kubasik~M, Koschorreck~M, Crepaz~H,
  Eschner~J, Polzik~E~S 2005
  Conditions for spin squeezing in a cold $^{87}$Rb ensemble
  \textit{J. Opt. B}  \textbf{7} S548

\bibitem{Windpassinger2007}
Windpassinger~P~J, Oblak~D, Petrov~P~G, Kubasik~M, Saffman~M,
  Garrido~Alzar~C~L, Appel~J, M\"uller~J-H, Kj{\ae}rgaard~N,
  Polzik~E~S 2008
   Nondestructive probing of Rabi oscillations on the cesium clock transition near the standard quantum limit
  \textit{Phys. Rev. Lett.} \textbf{100} 103601

\bibitem{Hilliard2008}
Hilliard~A, Kaminski~F, Le~Targat~R, Olausson~C, Polzik~E~S,
 Müller~J-H 2008
 Rayleigh superradiance and dynamic Bragg gratings in an end-pumped Bose-Einstein condensate
  \textit{Phys. Rev. A } \textbf{78} 051403

\bibitem{Hansen2001}
Hansen~H, Aichele~T, Hettich~C, Lodahl~P, Lvovsky~A, Mlynek~J,
  Schiller~S 2001
  Ultrasensitive pulsed, balanced homodyne detector: application to time-domain quantum measurements
   \textit{Opt. Lett. } \textbf{26} 1714

\bibitem{Appel2008}
Appel~J, Windpassinger~P~J, Oblak~D, Busk~Hoff~U, Kjaergaard~N,
  Polzik~E~S 2008
  Mesoscopic atomic entanglement for precision measurements beyond the standard quantum limit
  \textit{preprint arXiv:}0810.3545

\bibitem{CSA-Noise}
Bertuccio~G, Pullio~A 1993
 A method for the determination of the noise parameters in preamplifying systems for semiconductor radiation
detectors
 \textit{Rev.Sci.Instr.} \textbf{64} 3294

\bibitem{Scully1997}
Scully~M~O, Zubairy~M~S \textit{Quantum Optics} (Cambridge
University Press, 1997)

\bibitem{Reif1987}
Reif~F S, \emph{Fundamentals of Statistical and Thermal Physics}
(McGraw-Hill Book Company, 1965)

\bibitem{Barnes1971}
Barnes~J, { \it et al.} 1971
 Characterization of frequency stability
 \textit{IEEE transactions on instrumentation and measurement} \textbf{20} 105



\bibitem{Allan1966}
Allan~D~W 1966
 Statistics of atomic frequency standards
 \textit{Proceedings of the IEEE} \textbf{54} 221

\end{thebibliography}
\end{document}